\documentclass[onecolumn,referee,usenatbib]{mn2e}
\usepackage{graphicx}

\usepackage{amsmath,amssymb,amsfonts,epsfig}

\newcommand{\beqn}{\begin{eqnarray}}
\newcommand{\eeqn}{\end{eqnarray}}
\newcommand{\nbn}{\nabla_{\nu}}

\newcommand{\sbs}{\begin{subequations}}
\newcommand{\sbe}{\end{subequations}}

\newcommand{\beq}{\begin{equation}}
\newcommand{\eeq}{\end{equation}}

\def\0{{\bf 0}}
\def\1{{\bf 1}}
\def\2{{\bf 2}}
\def\3{{\bf 3}}

\begin{document}

\title{The traditional approximation in general relativity}

\author[A. Maniopoulou \& N. Andersson]
{Asimina Maniopoulou and Nils Andersson\\
Department of Mathematics, University of Southampton, Southampton
SO17 1BJ, UK}

\date{\today}

\maketitle

\begin{abstract}
We discuss the generalisation of the so-called traditional approximation, 
well known in geophysics, to general relativity. We show that the 
approximation 
is applicable for rotating relativistic stars provided that one 
focuses on relatively
thin radial shells. This means that the framework can be used to study 
waves in neutron star oceans. We demonstrate that, once the effects 
of the relativistic frame-dragging are accounted for, the angular 
problem reduces to Laplace's tidal equation. We  derive the 
dispersion relation for various classes of 
waves in a neutron star ocean and show that 
the combined effects of the frame-dragging and the gravitational 
redshift typically lower the frequency of a mode by 
about 20\%.
\end{abstract}

\section{Motivation}

Accreting neutron stars exhibit a variety of quasiperiodic phenomena, with frequencies
ranging from a few Hz to well above 1~kHz. Many of the observed oscillations are 
likely associated with 
processes in the accretion disk, but it is also probable that some 
of the observed features are linked to waves in the neutron star ocean \citep{bc,heyl,stroh}. 
For example, it has been observed that neutron stars that accrete matter at rates 
$\gtrsim 10^{-9} M_{\odot} \mathrm{yr}^{-1} $(the  so-called Z-sources) show quasi-periodic 
oscillations with frequencies of $\sim 6$ Hz. It has also been argued that the Z-sources 
are covered by massive oceans comprising a mass of $\sim 10^{-6} M_{\odot}$ 
composed of a degenerate liquid of light elements (C, O, Ne, Mg,..). 
This led \citet{bc} to argue that the observed oscillations may correspond
to ocean g-modes.

To analyse this suggestion in more detail \citet{buc} investigated the nature of 
low-frequency waves in the ocean of a rotating neutron star. They considered 
slowly rotating stars (far below the breakup limit) in Newtonian gravity, 
and made use of the assumption that 
$\Omega\gg\sigma$, where $\Omega$ is the rotation  frequency 
and $\sigma$ is the mode frequency. This is a reasonable assumption since
neutron stars accreting at a high rate are expected to spin much faster than a few Hz.
The studies of \citet{bc} and \citet{buc} both made use of the so-called
``traditional'' approximation, which 
has its origins in geophysics. 
This approximation leads to a separation of variables 
for the linearised Euler equations, where the angular problem is described in terms of 
Laplace's tidal equation. The solutions to this equation have been discussed in detail 
by many authors \citep{mkb,mil,buc,ls,town}.

The purpose of this paper is to extend the 
``traditional approximation'' to general relativity. The main motivation for this is that 
we know from the studies of \citet{bc,buc} \citet{town} that the
approximation provides a relatively simple way of analysing 
ocean waves in rotating Newtonian stars.
Yet there are several reasons for studying this problem in the framework of general relativity.
First of all, neutron stars are highly relativistic objects with $GM/Rc^2\approx 0.15-0.2$. 
This means that one must account for the gravitational redshift when comparing observed 
oscillations to theoretical models. A second relativistic effect that needs to be accounted for 
is the rotational frame-dragging. Finally, the traditional approximation scheme is 
interesting from a conceptual point of  view. Especially since it provides an 
alternative approach to existing work on the low-frequency
oscillations of rotating relativistic stars (see \citet{lock} for a discussion
and references to the relevant literature). 

\section{The relativistic equations of motion}

We begin by deriving the equations that govern a small amplitude ocean wave 
on a relativistic star. 
We consider a star which rotates uniformly at frequency 
$\Omega\ll (GM/R^3)^{1/2}$, where $M$ and $R$ are the neutron star mass and radius.
This means that the star rotates at a small fraction of the break-up 
velocity, allowing us to  neglect the ``centrifugal force'' 
and treat the unperturbed star as spherical. 
This approximation is justified for most observed neutron stars. However, it
is imperative to account for the  Coriolis force. In particular since it
leads to the existence of a new class of oscillation modes, the so-called
``inertial modes'' \citep{provost,lock}.

The spacetime of a slowly rotating star is described by the metric
\begin{equation}
ds^2 = - e^\nu dt^2 + e^\lambda dr^2 -2\omega r^2 \sin^2 \theta dt \, d\varphi 
+r^2(d\theta^2 + \sin^2 \theta d\varphi^2),
\end{equation}
where $\omega$ represents the dragging of inertial frames \citep{hartle}. Physically, 
 $\omega$ represents the angular velocity of a ``zero-angular momentum observer''
(for which the associated specific
angular momentum, $u_{\varphi}$, vanishes)
relative to infinity.
For an observer  co-moving  with the fluid, the four-velocity  
moves on wordlines of constant $t$ and $\theta$, and the corresponding 
 angular velocity $\Omega$  is defined as the ratio 
\beq 
\Omega=\frac{u^{\varphi}}{u^{t}}=\frac{d\varphi}{dt} \label{angv}.
\eeq   
Combining this with the requirement that 
\beq 
u^{\mu}\,u_{\mu}=-1 \label{nv}
\eeq
we have
\begin{equation}
u^\mu = [ e^{-\nu/2}, 0, 0, \Omega e^{-\nu/2} ].
\end{equation}

We wish to consider small perturbations of the star. 
Since we are mainly interested in ocean waves we will
use the Cowling approximation; i.e. assume that all 
metric perturbations can be neglected. This is a reasonable assumption given 
the relatively low density of a neutron star ocean. One would, for example, 
not expect the ocean waves to radiate appreciable amounts of gravitational waves. 
We also need to neglect the dynamic nature of spacetime if 
we want to implement the ``traditional approximation'' in general relativity. 

In order to derive the equations that govern the fluid motions we assume that the 
stellar background is appropriately described by a perfect fluid equation of state 
with $p=p(\rho)$, which means that the energy-momentum tensor takes the form 
 \beq  
T^{\mu \nu}=\rho u^{\mu}u^{\nu}+p q^{\mu \nu},
\label{enermom}\eeq
where $ q^{\mu \nu} $ is the projection orthogonal to $u^{\mu}$; 
\beq 
q^{\mu \nu}=g^{\mu \nu}+u^{\mu}u^{\nu} \label{proj}
\eeq
and the scalars $\rho$ and $p$  denote the energy density and  pressure, 
as measured by a co-moving observer (an observer with four-velocity $u^{\mu}$),
respectively.

The equations of motion follow from the conservation law
\beq 
\nabla_{\mu}T^{\mu\, \nu}=0 \label{cont},
\eeq
as well as the  Einstein field equations. It is well-known that, in the case of slow-rotation, this
problem reduces to the standard Tolman-Oppenheimer-Volkoff equations for a non-rotating star,
plus a single ordinary differential equation determining the frame-dragging \citep{hartle}. 
These equations can be written 
\begin{eqnarray}
\lambda^\prime &=& - {2 e^\lambda (M-4 \pi \rho r^3) \over r^2}, \\
\nu^\prime &=&  {2 e^\lambda (M+4 \pi p r^3) \over r^2 }\\
p^\prime &=& -{p+\rho \over 2} \nu^\prime,
\end{eqnarray}
where  primes denote  derivatives with respect to $r$, and $M=M(r)$ is
the ``mass within radius $r$'', which follows from the relation 
\begin{equation}
e^\lambda = \left( 1 - {2M \over r} \right)^{-1} \ .
\end{equation}
To incorporate the leading order rotational effects, we  need to solve an 
additional equation that describes the frame-dragging, $\omega$. 
It is  convenient to use 
the difference $\tilde{\omega} = \Omega -
\omega$,  corresponding to the rotation frequency as measured
by local zero-angular momentum observers, which is determined by the equation
\begin{equation}
 \left[ e^{-(\nu + \lambda)/2} \tilde{\omega}^\prime
\right]^\prime + {4 \over r}  e^{-(\nu + \lambda)/2}  \tilde{\omega}^\prime
-16 \pi (p+\rho) e^{(\lambda - \nu)/2} \tilde{\omega} = 0 \ ,
\label{framed}\end{equation}
We also need boundary conditions for $\tilde{\omega}$ 
at the origin and spatial infinity. At the origin we require that 
$\tilde{\omega}$ is regular.
In the vacuum outside the star, equation (\ref{framed}) can be solved analytically, and
we have
\begin{equation}
\tilde{\omega} = \Omega - {2J \over r^3},
\end{equation}
where $J$ is the total angular momentum of the star.
This relation can  be used to provide boundary conditions for 
$\tilde{\omega}$ (and its derivative) at the surface of the star in terms
of $\Omega$ and $J$.
Specifically, the solution to equation (\ref{framed}) is normalised by requiring that
\begin{equation}
\tilde{\omega}(R) + {R\tilde{\omega}^\prime(R)\over 3} = \Omega \ .
\label{dragging}\end{equation}

In order to derive the equations that govern perturbations, it is useful to 
work with 
projections of equation (\ref{cont}) along, and orthogonal to, the four-velocity,
cf. \citet{ipser}. Projecting along  $u^\mu$ we have the 
relativistic continuity equation
\begin{eqnarray}
u^{\mu} \nabla_{\mu} \rho + (p + \rho) \nabla_\mu u^{\mu} = 0. 
\label{along}\end{eqnarray}
The orthogonal projection leads to the 
relativistic Euler equations
\beqn
(p+\rho) u^\mu \nabla_\mu u_\nu + q^\mu_\nu \nabla_\mu p = 0. 
\label{ortho}\eeqn
Perturbing equation (\ref{along}), assuming that the Cowling approximation is made
(i.e. taking $\delta g_{\mu \nu}=0$),
we get
\beqn 
\delta u^{\mu} \nabla_\mu \rho +u^{\mu} \nabla_\mu \delta\rho +(\rho + p)\nabla_\mu 
\delta u^{\mu} = 0. 
\label{asta} \eeqn
Meanwhile, the perturbed  Euler equations can be written as
\begin{eqnarray}
 (\delta \rho + \delta p) u^{\mu}\nabla_\mu u^{\kappa} &+& (p+\rho) \delta u^{\mu}\nabla_\mu u^{\kappa} + (g^{\nu \kappa}+u^{\nu}u^{\kappa}) \nbn \delta p + (p+\rho) u^{\mu}\nabla_\mu \delta u^{\kappa} \nonumber \\
&-&\delta u^{\nu}u^{\mu}u^{\kappa} (p+\rho) \nabla_{\mu}u_{\nu}=0. \label{dike}
\end{eqnarray}

The perturbed fluid velocity has
 (quite generally) components
\begin{eqnarray}
\delta u^t &=&  (\Omega-\omega) r^2 \sin^2 \theta e^{-\nu/2} 
H_2 e^{i(\sigma t+ m\varphi)},\\
\delta u^r &=& e^{\nu/2} W e^{i(\sigma t+ m\varphi)},\\
\delta u^\theta &=& e^{\nu/2} H_1 e^{i(\sigma t+ m\varphi)}, \\
\delta u^\varphi &=& e^{\nu/2} H_2 e^{i(\sigma t+ m\varphi)},
\end{eqnarray}
where $\delta u^t$ has been fixed by the requirement that 
the overall four-velocity $u^\mu + \delta u^\mu$ is normalised in the standard way. 
Here $W$, $H_1$ and $H_2$ are functions of  $r$ and $\theta$, and
$\sigma$ is the oscillation frequency observed in the inertial frame.
In addition we let 
\begin{eqnarray}
\delta p  &=&  p_1 e^{i(\sigma t + m\varphi)}, \\
\delta \rho &=&  \rho_1 e^{i(\sigma t + m\varphi)}. 
\end{eqnarray}
where $p_1$ and $\rho_1$ are functions of  $r$ and $\theta$.
Given these definitions, the continuity equation  (\ref{asta}) 
leads to 
\begin{eqnarray}
i(\sigma + m\Omega) \rho_1 &+& (p+\rho) e^\nu W^\prime +
\left[ (p+\rho) \left( {\lambda^\prime \over 2} + \nu^\prime + {2 \over r}\right) + \rho^\prime \right]e^\nu W \nonumber \\
&+& (p+\rho) e^\nu 
\left[ {\partial H_1 \over \partial \theta} + {\cos \theta \over
\sin \theta} H_1 \right] + im (p+\rho) e^\nu H_2 = 0 
\label{con1}\end{eqnarray}
The $r$-component of (\ref{dike}) gives
\begin{eqnarray}
&&
2\,r\left (\rho+p\right )\left[ r\left (\omega-\Omega\right )
\nu^\prime-\omega^\prime r+2\,
(\Omega-\,\omega)\right] \sin^2 \theta H_2
 \nonumber \\
&& - 2\, p_1^\prime
 - \left ( p_1 + \rho_1 \right )\nu^\prime
-2i(\sigma +m\Omega)\,\left (\rho+p\right ){e^{\lambda}}W = 0,
\label{r_one} \end{eqnarray}
while the $\theta$ and $\varphi$ components lead to, respectively,
\begin{eqnarray}
&&
2\,\left (\rho+p\right ){r}^{2} \left (\Omega-\omega\right )
\cos \theta \sin \theta H_2 -{\frac {\partial  p_1 }{\partial \theta}}
 - i(\sigma+m\Omega)\,\left (\rho+p\right ){
r}^{2}H_1
= 0,
\label{th_one}\end{eqnarray}
and
\begin{eqnarray}
2(p+\rho)r^2 (\Omega-\omega) \sin \theta \cos\theta H_1 &+& im  p_1
+i (\sigma + m\Omega) (p+\rho) r^2 \sin^2 \theta H_2 \nonumber \\
&+& (p+\rho) [ (\Omega-\omega)(2r-r^2\nu^\prime)-r^2 \omega^\prime ]
\sin^2\theta W = 0.
\label{ph_two}\end{eqnarray}

It is worth emphasising the following point:
In deriving
the above equations we have neglected terms proportional to $\sigma \Omega$ in 
the $\varphi$-component of the continuity equation (\ref{dike}). 
Such an approximation is not required in the Newtonian case, 
as the terms that we have omitted are post-Newtonian contributions. 
Specifically, the complete 
$\varphi$-component of equation (\ref{dike}) includes the 
term
$$
i\dfrac{\sigma(\Omega-\omega)}{c^2}r^2\sin^2\theta p_1
$$
where we have reinstated the speed of light ($c$).
This term vanishes in the Newtonian limit ($c\rightarrow\infty)$, but in the relativistic 
case we can neglect it only when $\sigma$ is suitably small.  

We do not, however, expect this difference between the Newtonian and the relativistic 
equations to have great effect on the ocean waves that we are interested in. 
The main reason for this is 
that: i) the Newtonian results show that the 
g-modes of the ocean have frequencies that scale as $\sqrt{\Omega}$ \citep{buc}, and
ii) inertial modes, like the r-modes, would have frequencies of order 
$\Omega$ \citep{lock}. In other words, at least for these classes of modes,
our equations should be consistent.

\section{The traditional approximation}

We now wish to analyse the various oscillation modes of a neutron star ocean.
To do this we follow e.g. \citet{buc} and
introduce the traditional approximation. As in the Newtonian problem, this 
 corresponds to 
\begin{itemize}
\item neglecting the Coriolis term in the $r$-momentum equation,
\item assuming that the mode is mainly horisontal.
\end{itemize}
In our case, the first assumption corresponds to neglecting the $H_2$ term in 
equation (\ref{r_one}), while the second implies that we should neglect
$W$ in equation (\ref{ph_two}).

With these simplifications equations (\ref{th_one}) and (\ref{ph_two})
can be solved for $H_1$ and $H_2$. This then gives us
\begin{equation}
-i (\sigma+m\Omega) (p+\rho)r^2 \sin \theta ( 1 - q^2 \cos^2 \theta) H_1 = 
  \sin \theta {\partial p_1 \over \partial \theta }+  
m q \cos \theta p_1
\end{equation}
and
\begin{equation}
-  (\sigma+m\Omega) (p+\rho)r^2 \sin \theta ( 1 - q^2 \cos^2 \theta) H_2 =
2 q \sin \theta \cos\theta 
{\partial p_1 \over \partial \theta} +  m  p_1 
\end{equation}
Here we have introduced 
\begin{equation}
q = { 2(\Omega -\omega) \over \sigma + m\Omega} \ .
\end{equation}

By using the above results in (\ref{con1}) we find
\begin{equation}
 i(\sigma+m\Omega) \rho_1 + (p+\rho) e^\nu W^\prime +
\left[ (p+\rho) \left( {\lambda^\prime \over 2} + \nu^\prime + {2\over r}
\right) + \rho^\prime \right] e^\nu W = - {i e^\nu \over (\sigma + m\Omega) r^2} L (p_1), 
\label{eq1}\end{equation}
where we have defined the operator
\begin{equation}
L = {\partial \over \partial \mu } \left( { 1 - \mu^2 \over 1
- q^2 \mu^2} { \partial \over \partial \mu} \right) - 
{ m^2 \over (1-\mu^2) (1 - q^2 \mu^2)} - {mq (1 +q^2 \mu^2) \over
(1-q^2 \mu^2)^2},
\end{equation}
with $\mu = \cos \theta$. Notably, this has exactly the same form as
the angular operator in the Newtonian problem, cf. equation (7) in \citet{buc}. 
One has to be careful 
here, though, because in the relativistic case $L$ depends on $r$ 
through $q$ (which in turn depends on $\omega(r)$). 
However, as long as we are interested in 
oscillations in a thin radial shell we may take the frame-dragging to
be a constant. 

Aiming to write the final equations in a form reminiscent of the 
Newtonian results, we relate the
displacement component $\xi^r$  to the perturbed velocity
component via
\begin{equation}
\delta u^\mu = q^\mu_{\ \nu} \pounds_u \xi^\nu,
\end{equation}
where the projection $q^{\mu\nu}$ was defined earlier, and where 
$\pounds_u$ is the Lie derivative along $u^\mu$ 
This means that
\beqn
\xi ^{r} &=& \dfrac{e^{\nu}\,W(r,\theta)e^{i(\sigma t+m\phi)}}{i(\sigma +m\Omega)}, \\
\xi ^ {\theta}&=&\dfrac{e^{\nu}\,H_1(r,\theta)e^{i(\sigma t+m\phi)}}{i(\sigma +m\Omega)}, \\
\xi ^{\phi} &=&\dfrac{e^{\nu}\,H_2(r,\theta)e^{i(\sigma t+m\phi)}}{i(\sigma +m\Omega)}.
\eeqn
We have used the inherent gauge freedom to set the  temporal displacement to zero, i.e. 
$\xi^{t}=0$.
We also  relate the Eulerian perturbations in density and pressure
in the standard way;
\beqn
{\Delta p \over \Gamma_1 p } = {\Delta \rho_1 \over p + \rho} \Rightarrow
\dfrac{p_{1}+\xi ^{\mu}p_{\,;\,\mu}}{\Gamma_1 p}= \dfrac{\rho_{1}+  \xi^{\mu}\rho_{\,;\,\mu}}{\rho +p}\Rightarrow 
 \rho_1 = {p + \rho \over \Gamma_1 p} p_1 -\xi^{r} 
\left[ \rho^\prime - {p + \rho \over \Gamma_1 p} p^\prime \right], 
\label{eulp}\eeqn
where $\Gamma_1$ is the adiabatic exponent 
$\Gamma_1=\left(\dfrac{\partial \ln p}{\partial \ln \rho }\right)$ evaluated at constant 
entropy.

Given these various relations we can write equation
(\ref{eq1}) as
\begin{equation}
-  \xi^{r\prime} -
\left[ { p_1 \over \Gamma_1 p} + \left( {p^\prime \over
\Gamma_1 p} + {\lambda^\prime \over 2} + {2\over r} 
\right) \xi^r \right] = {p e^{\nu} \over (\sigma+m\Omega)^2 r^2 (p+\rho)} 
L\left( {p_1 \over p} \right) 
\label{eq1b}
\end{equation}

Finally, 
the $r$-momentum equation (\ref{r_one}) reads (after neglecting the Coriolis term) :
\begin{equation}
(p+\rho) e^{\lambda-\nu}\left[ (\sigma + m \Omega)^2 + {\nu^\prime e^{\nu-
\lambda/2} \over 2} \mathcal{A} \right]
\xi^r =   p_1^\prime + {\nu^\prime \over 2\Gamma_1 p}\left[\rho+(1+\Gamma_1)p\right] p_1,
\label{eq2}
\end{equation}
where we have defined the 
relativistic Schwarzschild discriminant as
\begin{equation}
\mathcal{A} = {e^{-\lambda/2} \over p+\rho} \left( \rho^\prime
- {p+\rho \over \Gamma_1 p} p^\prime \right).
\end{equation}
For later convenience, we also introduce the relativistic analogue of the Brunt-V\"ais\"al\"a 
frequency as
\begin{equation}
 N^2 = - \dfrac{\nu^\prime}{2}\mathcal{A}
\end{equation}.

\section{The Newtonian limit}

As a consistency check of our derivation we want to compare our final formulas
(\ref{eq1b}) and (\ref{eq2}) to the corresponding results in the Newtonian case. 
To facilitate this comparison we need to transform our expressions, which are determined in
a coordinate basis, into an orthonormal basis. This basis should be the relativistic 
analogue  of the flat space-time orthonormal basis used by \citet{buc};
\beq 
\vec{\xi}_\mathrm{N}=\xi^{r}_\mathrm{N}\hat{r}+\xi^\theta_\mathrm{N}\hat{\theta}+
\xi^\varphi_\mathrm{N}\hat{\varphi},
\eeq
where the subscript $\mathrm{N}$ denotes a Newtonian vector.  
 
In the corresponding relativistic basis, the spatial components $\xi^{a}_\mathrm{o}$
of the displacement vector are determined from
\beqn \xi^{r}_\mathrm{o}&=&e^{\lambda /2}\xi^{r}\ , \\
\xi^{\theta }_\mathrm{o}&=&r\xi^{\theta}\ ,\\ 
\xi^{\phi}_\mathrm{o}&=&r \sin\theta \xi^{\phi}\ ,
\eeqn
In this frame we can  identify the components of the displacement  with their   
Newtonian counterparts. 
Moreover, we can argue in favour of the validity of the traditional approximation 
in a way analogous to that of \citet{buc}.

In the orthonormal frame equation (\ref{r_one}) becomes
\beqn 
e^{-\nu+\lambda/2}\left[(\sigma+m\Omega)^2+\dfrac {\nu^\prime e^{\nu-\lambda/2}}{2}\mathcal{A} \right]
\xi^{r}_\mathrm{o}=p_1^\prime + {\nu^\prime \over 2}
\left(\dfrac{\rho + p}{p\Gamma_{1}}+1\right) p_1\nonumber\\
+ i e^{-\nu}(\sigma+m\Omega)\left[(\Omega-\omega)
\left(r\nu^\prime +2\right)+r\omega^\prime \right]\sin\theta \xi^{\varphi}_\mathrm{o}. \label{prin3}
\eeqn
We can safely neglect the Coriolis term (the last term of the right-hand side) as long as 
\beqn 
\left| N^2\xi^{r}_\mathrm{o}\right| \gg e^{- \nu} (\sigma+m\Omega) \left|
 \left[(\Omega-\omega)\left(r\nu^\prime +2\right)+r\omega^\prime \right]
\sin\theta \xi^{\varphi}_\mathrm{o}\right|\ ,
\eeqn
or equivalently
\beqn \left| N^2\right| \gg (\sigma+m\Omega)\left| \left[(\Omega-\omega)
\left(r\nu^\prime +2\right)+ r\omega^\prime \right] \right| 
\left| \dfrac{\xi^{\varphi}_\mathrm{o}}{\xi^{r}_\mathrm{o}} \right| .\label{approx1}
\eeqn
Since we are assuming that the mode is horisontal, it is far from obvious that this 
inequality will hold. However, for a neutron star ocean the order of magnitude
of $r\omega^\prime $ can be estimated from the value at the surface. From 
equation (\ref{dragging}), we see that
\beq
r\omega^\prime\approx R \tilde{\omega}^\prime (R) = 3\tilde{\omega}(R)\approx \dfrac{6I\Omega}{R^3}
\approx \dfrac{12 M\Omega}{5R}
\eeq
and since  $M/R\approx 1/5$ for a typical neutron star it follows that 
$r\omega^\prime\approx \dfrac{12}{25}\Omega$. In a similar way, 
\beq
\nu^\prime  \approx {2e^\lambda M \over r^2} \approx { 2 M \over R^2}
\eeq
which means that  $r\nu^\prime\approx 2/5$. Combining the various terms the 
inequality becomes
\beqn 
\left| N^2\right| \gg 1.5 \sigma \Omega  \left| \dfrac{\xi^{\varphi}}{\xi^{r}_\mathrm{o}}\ \right| . 
\label{approx2}\eeqn
According to \citet{bc}, the thermal buoyancy in the deep ocean leads to 
$N\sim 1-5$~kHz. Using the slow-rotation solutions one can also argue that
$| \xi^{\varphi}_\mathrm{o} / \xi^{r}_\mathrm{o}|  \sim 
10^4\mbox{m} /10^2\mbox{m}=10^2$. This means that, for mode frequencies below (say) 
 $10$~Hz  our approximation remains valid for rotation frequencies up to a few hundred Hz. 
Given that the highest observed spin rates in Low-Mass X-ray Binaries are about 600~Hz,  
our formulas are likely to be useful for studies of low frequency ocean waves in most
accreting neutron stars. 

In the orthonormal 
frame our two final equations are
\beqn 
e^{-\nu+\lambda/2}\left[(\sigma+m\Omega)^2+\dfrac {\nu^\prime e^{\nu-\lambda/2}}{2}\mathcal{A} \right]
\xi^{r}_\mathrm{o}=
p_1^\prime + {\nu^\prime \over 2}\left(\dfrac{\rho + p}{p\Gamma_{1}}+1\right) p_1
\label{final1}. \eeqn
and
\beqn 
\xi^{r\prime}_\mathrm{o}+\left(\dfrac{p^\prime}{p \Gamma_1}+\dfrac{2}{r}\right)\xi^{r}_\mathrm{o}+
\dfrac{e^{\lambda/2}p_1}{p\Gamma_1}=- \dfrac{e^{\nu+\lambda/2}p}{(\rho+p)(\sigma+m\Omega)^2 r^2}
L\left(\dfrac{p_1}{p}\right).
\label{final2}\eeqn
We want to  verify that these equations limit to the expected Newtonian ones.
To check this we note that the Newtonian limit corresponds to
\begin{eqnarray*}
\vert \Phi \vert &\ll&  1 \mbox{ where } \, \Phi=-M/r \, \mbox{ is the Newtonian gravitational potential ,}
\\
p  & \ll & \rho \\
\omega &\to & 0 \\
e^{-\lambda} &\approx & e^{\nu} = 1 - {2M \over r} \approx 1+2\Phi \approx 1 \\
\nu^\prime &=& 2\Phi^\prime\approx {2M \over r^2} = 2g
\end{eqnarray*} 
where $g$ is the gravitational acceleration. Using these simplifications, letting
$\xi^j_\mathrm{o} \to \xi^j_\mathrm{N}$, 
as well as introducing the scale-height $h=p/\rho g$,  
 equation (\ref{final1}) yields equation (5) of \citet{buc};
\beqn 
\rho(\sigma^2_r -N^2)\xi_\mathrm{N}^{r}=p_1^\prime +\dfrac{p}{h \Gamma_1}\label{buc5},
\eeqn
where $\sigma_r=\sigma+m\Omega$ 
is the mode frequency in the rotating frame. At the same time, 
equation (\ref{final2}) reads 
\beqn 
-\xi_\mathrm{N}^{r\prime}-\dfrac{p_1}{\Gamma_1 p}+\left(\dfrac{\rho g}{p\Gamma_1}-\dfrac{2}{r}\right)\xi_\mathrm{N}^{r}=
\dfrac{hg}{\sigma^2_R r^2}L\left(\dfrac{p_1}{p}\right)
\label{buc6a}.
\eeqn
which is 
equation (6) of \citet{buc} (where the term  $2\xi_N^r/r$ was omitted due to a misprint).

\section{The ocean g-modes}

We now want to assess the magnitude of the relativistic effects on 
the ocean g-modes of a rotating neutron star. In principle, we could do this
by solving  (\ref{final1}) and (\ref{final2}) numerically, thus finding the modes of 
oscillation for a given model of the ocean. It may, however, be more instructive to 
extract the required information through an approximate analytic calculation. 
As this is the route we have chosen, we 
 first simplify our equations somewhat by assuming that
\begin{eqnarray}
p &<< & \rho, \\
e^{-\lambda} &=&  e^\nu \approx 1 - {2M \over r}, \\
\nu^\prime  &\approx& {2e^\lambda M \over r^2} \equiv 2 e^{\lambda/2} g, \\
\omega &=& {2J \over r^3}
\end{eqnarray}
These approximations should be quite accurate in the low-density ocean 
of a neutron star. 
We also denote the ``angular eigenvalues'' by $\alpha$, i.e. we assume that
\begin{equation}
L \left( {p_1 \over p} \right) = -\alpha  
\left( {p_1 \over p} \right).
\end{equation}

\begin{figure}
\centerline{
\includegraphics[width=10cm,clip]{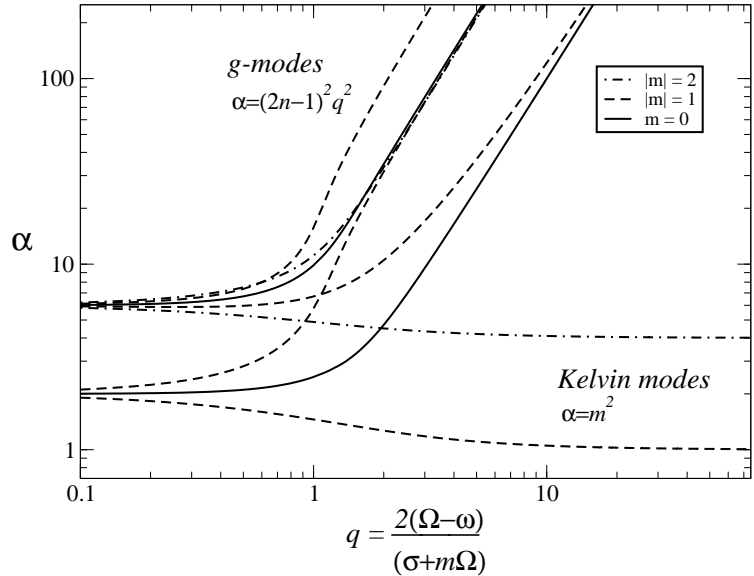}}
\caption{An illustration of the relation between the numerically determined 
angular eigenvalues 
$\alpha$ and the spin parameter $q$.	
The eigenvalues for the g-modes asymptote to 
$\alpha\approx (2n-1)^2q^2$ for $n=1,2,3...$ 
while the trapped Kelvin modes correspond to $\alpha\approx m^2$ for large $q$. 
 }
\label{buca}
\end{figure}

In Figure~\ref{buca} we show a sample of numerically determined 
angular eigenvalues $\alpha$. These results were calculated using a 
method similar to that employed by \citet{buc}. These results clearly 
illustrate the large-$q$ asymptotic behaviour. 
It should be noted that the eigenvalues can be classified in different 
ways, cf. \citet{buc,ls}. For example, in the $q\to0$ limit we can 
associate each solution with the $l$ and $m$ of the single 
spherical harmonic eigenfunction. Note that our data 
corresponds to $l=1$ and $l=2$ and the $2l+1$ solutions 
that correspond to the various permitted values of $m$ are 
easily distinguished in the figure. Alternatively we could label the
modes by their 
symmetry with respect to the equator (it depends on whether $l+m$ is 
odd or even). It should be noted that, for large $q$, the first three g-mode asymptotes
then correspond to odd, even and odd eigenfunctions, respectively.     

Having defined the angular eigenvalue $\alpha$, 
we obtain the  two equations
\begin{equation}
-  \xi^{r\prime }_\mathrm{o} -
\left\{ \left( 1 - {2M \over r} \right)^{-1/2}
{ p_1 \over \Gamma_1 p} - \left[ \left( 1 - {2M \over r} \right)^{-1/2}
{ \rho g \over
\Gamma_1 p}  - {2\over r} 
\right] \xi^{r}_\mathrm{o} \right\} 
= - \left( 1 - {2M \over r} \right)^{-1/2} {\alpha  p_1 \over \rho (\sigma+m\Omega)^2 r^2 } 
\end{equation}
and
\begin{equation}
\rho \left[\left( 1 - {2M \over r} \right)^{-3/2}  (\sigma + m \Omega)^2 +  
\left( 1 - {2M \over r} \right)^{-1/2} g  \mathcal{A}\right] 
\xi^{r}_\mathrm{o} =   p_1^\prime         
\left( 1 - {2M \over r} \right)^{-1/2} {\rho g \over \Gamma_1 p} p_1.
\end{equation}

These equations simplify considerably if we introduce new variables
$\eta = r^2 \psi\xi^r$ and $\phi = p_1 /\psi$, 
where
\begin{equation}
\psi =  \exp \left[ - \int \left( 1 - {2M \over r}\right)^{-1/2} {g \over c_s^2} dr \right]
\end{equation}
and the ``sound speed'' follows from
\begin{equation}
c_s^2 =  {p \Gamma_1 \over \rho} 
\end{equation}
In terms of these new variables we get
\begin{equation}
- \eta^\prime = \left( 1- {2M \over r} \right)^{-1/2}
\left[ { 1 \over \Gamma_1 p} - \left( 1- {2M \over r} \right) { \alpha \over 
(\sigma + m\Omega)^2 r^2 \rho} \right] r^2 \phi  \psi^2
\end{equation}
and
\begin{equation}
\phi^\prime =  
\left( 1- {2M \over r} \right)^{-1/2} \left[ (\sigma + m\Omega)^2 +  
\left( 1- {2M \over r} \right) gA \right] \dfrac{\rho  \eta}{r^2 \psi^2}
\end{equation}

If we now assume that the radial 
dependence of the perturbations is described by $\eta \sim \phi \sim \exp(ik r)$, 
where $k$ is the radial wavenumber,   we readily 
arrive at the dispersion relation
\begin{equation}
k^2 =  \left( 1- {2M \over r} \right)^{-1}  \left[ (\sigma + m\Omega)^2 +  
\left( 1- {2M \over r} \right) g\mathcal{A} \right]
\left[ { \rho \over \Gamma_1 p} - \left( 1- {2M \over r} \right) 
{ \alpha \over 
(\sigma + m\Omega)^2 r^2 } \right]
\end{equation}
We now use the   Brunt-V\"ais\"al\"a frequency $N^2 = -g\mathcal{A}$ and 
an analogue to the standard Lamb frequency;
\begin{equation}
L^2_\alpha = {\alpha c_s^2 \over r^2}
\end{equation}
Implementing a dimensionless corotating oscillation frequency 
\beq
\kappa  = \dfrac{\sigma + m\Omega}{\Omega}
\eeq
we  get the final result
\begin{equation}
k^2 = { \Omega^2 \over c_s^2 \kappa^2} \left( 1- {2M \over r} \right)^{-1}
\left[\kappa^2 - \left( 1- {2M \over r} \right) \left( {N\over \Omega} \right)^2 
\right] 
\left[\kappa^2 - \left( 1- {2M \over r} \right) \left({ L_\alpha \over \Omega }\right)^2 \right].
\end{equation}

In order to estimate the g-mode frequencies we recall that
(cf. \citet{buc,town})
\begin{equation}
\alpha \approx (2n -1)^2 q^2 = (2n -1)^2 
\left[ {2 (1 -\omega/\Omega) \over \kappa} \right]^2 \qquad n=1,2,3,...
\label{kappa}\end{equation}
for large $q$ (see \cite{buc} for a discussion of the interpretation of the 
``quantum number'' $n$). 
This behaviour is apparent in Figure~\ref{buca}. 
Expanding equation (\ref{kappa}) for low
frequencies  (that is, for
$ \kappa << N/\Omega$ and $\kappa << L_\alpha/\Omega$)
we find
\begin{equation}
\kappa \approx \sqrt{2} (2n -1)^{1/2}  \left[ 1 - {\omega \over \Omega} 
\right]^{1/2}  
\left( 1- {2M \over r} \right)^{1/4} \left( {N \over k r \Omega} \right)^{1/2}.
\label{gfreq}\end{equation}
Even though this does not actually determine the mode-frequencies 
it allows us to estimate the magnitude of the relativistic effects.
We see that there are two main effects: 
The obvious one is the gravitational redshift factor, which 
for a typical neutron star with $M/R \approx 1/5$   
decreases the  frequency by about 10\% compared to the Newtonian case; 
\beq
\left( 1 - { 2M \over R} \right)^{1/4} \approx 1 - { M \over 2R} \approx 0.9.
\eeq
The rotational frame-dragging adds a further 10\% to the decrease
since
\begin{equation}
\omega \approx {2J \over R^3} = {2I\Omega \over R^3} \approx {4M\Omega \over 
5R} \quad \longrightarrow \quad 
\left[ 1 - {\omega \over \Omega} 
\right]^{1/2}   \approx 0.9.
\end{equation}
Thus, we would expect the relativistic ocean g-modes to have frequencies
roughly 20\% lower than their Newtonian counterparts.

It is easy to argue that these relativistic corrections must be included 
if one wants to draw the correct conclusions from observed oscillations.
 For example, comparing our relativistic result
to the corresponding Newtonian one (where $\omega$ and $M/R$ are 
both taken to be zero in (\ref{gfreq})) we see that already for $n>2$ would we get
$\kappa_\mathrm{GR} (n+1) < \kappa_\mathrm{Newton}(n)$. This means that
the Newtonian formula would  predict that the wrong 
mode was observed. This could be crucial if the aim is to study the 
inverse problem to put constraints on the physics of neutron star 
oceans. 

As is clear from Figure~\ref{buca} there is a distinct class of angular 
eigenvalues which do not exhibit the scaling discussed above. 
These solutions correspond to trapped Kelvin waves, and 
\citet{town} has shown that for large  $q$ one gets
\beq 
\alpha=m^2 \dfrac{2m q}{2mq+1} \approx m^2 
\eeq
Following the same procedure as for the g-modes, we now find that 
\beq 
\kappa \approx m \left(1-\dfrac{2M}{R}\right)^{1/2}\dfrac{N}{\Omega k r}. 
\label{kelvin}\eeq
The derivation of this relation is valid as long as the radial wavenumber $k$ 
is taken to be large, i.e. for short-wavelength waves. Otherwise we would not have
$ \kappa << N/\Omega$ and $\kappa << L_\alpha/\Omega$. Now
comparing to the Newtonian result \citep{town} one finds that the redshift 
 reduces the frequency by about 20\%, since 
\beq 
\left(1-\dfrac{2M}{R}\right)^{1/2}\approx 1-\dfrac{M}{R}\approx 0.8 \ . 
\eeq   
compared to the Newtonian case.

\section{Ocean inertial modes: the r-modes}

Given the relativistic version of the equations of motion it 
is interesting to ask whether we can estimate the frequencies
of r-modes in the ocean. 
As discussed by eg. \citet{town} the r-modes are not well 
represented by the traditional approximation. However, it is straightforward
to  follow \citet{provost} and assume that the following scalings apply 
(see also \citet{lock}):
\begin{eqnarray*}
W &\sim & \sigma \Omega \sim \Omega^3 , \\
H_1 &\sim & H_2 \sim \sigma \sim \Omega, \\
p_1 &\sim & \rho_1 \sim \Omega^2, 
\end{eqnarray*}
The  relativistic continuity equation then immediately reduces to
\begin{equation}
{\partial \over \partial\theta } ( \sin \theta H_1) + im \sin \theta H_2 = 0,
\end{equation}
and a second equation relating $H_1$ and $H_2$ can be obtained by combining
the $\theta$- and $\varphi$-momentum equations. This leads to
\begin{eqnarray}
&& 2i( \sigma + 2m\Omega - m\omega) \sin\theta \cos\theta H_2 + i
(\sigma + m\Omega) \sin^2 \theta {\partial H_2 \over \partial \theta} + 
\nonumber \\
&& [ m (\sigma + m\Omega) + 2(\Omega-\omega ) (\cos^2\theta - \sin^2\theta)]
H_1
+ 2(\Omega-\omega)\sin \theta \cos\theta {\partial H_1 \over \partial \theta} = 0.
\end{eqnarray}

After some straightforward algebraic manipulations  one can 
show that these two equations will be satisfied if
\begin{equation}
H_1 = {1 \over \sin \theta} P_{lm}(\cos \theta)
\end{equation}
and
\begin{equation}
\sigma = -m\Omega + {2m(\Omega -\omega) \over l(l+1)}.
\end{equation}
It should, of course, be noted that this derivation only works as long we can take
the frame-dragging, $\omega$, to be a constant, i.e. for oscillations in a thin shell.
Anyway, we conclude that the frequencies of the ocean r-modes are sensitive to 
the surface value of the frame-dragging. The size of the effect can be appreciated if we
take the ratio between the relativistic co-rotating frequency $\kappa_\mathrm{GR}$ and its
Newtonian counterpart $\kappa_\mathrm{Newton}$. This immediately leads to
\beq
{ \kappa_\mathrm{GR} \over \kappa_\mathrm{Newton}} = 1 - {\omega \over \Omega} 
\approx 1 - { 4M \over 5R} \approx 0.84
\eeq
Thus, the relativistic frame-dragging lowers the r-mode frequencies (in the rotating frame)
by about 15\%. This is in accordance with the results of e.g. \citet{lock} and \citet{rezzolla}.

Finally, it is worth pointing out that the calculation of  general inertial 
modes is much more involved since one must account for the coupling of 
many spherical harmonics, see  \citet{lock} for a recent discussion.
In particular, such a calculation may be needed for the fastest spinning neutron stars
since the g-modes will also be dominated by the Coriolis force
when $\Omega>>N$. 

\section{Concluding remarks}

We have shown how the so-called traditional approximation can be 
generalised to general relativity. 
This leads to a set of equations that ought to prove
valuable for future studies of waves in the oceans of rotating neutron stars.
The derivation of these equations required several 
assumptions in addition to those made in the Newtonian case. 
First of all, we have to make use of the Cowling approximation. 
That is, we neglect all metric perturbations. This may seem a 
drastic simplification, but it is a reasonable approximation
for waves in the relatively low-density neutron star ocean. 
After all, such waves are not expected to 
lead to appreciable gravitational-wave
emission. Secondly, we have to assume that 
the frame-dragging $\tilde{\omega}$ is constant, which 
restricts us to consider thin shells. Again, this should be 
a good approximation
for neutron star oceans. Finally, and perhaps most interestingly, we have to 
neglect some
terms of post-Newtonian order from the perturbed relativistic
Euler equations, an approximation which is valid for low-frequency 
oscillations.  
Having made these various assumptions we can separate the eigenfunctions
into a radial and an angular part. The angular eigenvalue problem 
is formally identical to the Newtonian one: 
it corresponds to solving Laplace's 
tidal equation, which now depends also on the dragging of inertial frames. 
It should be emphasised that, since the two problems are identical we can  
use the results for the Newtonian angular problem also in the general 
relativistic case.  

Our calculation prepares the ground for detailed calculations based on sophisticated
models of the neutron star ocean. Given the eigenvalues to Laplace's tidal 
equation, we need to solve the two radial differential 
equations (\ref{final1}) and (\ref{final2}), together with the 
relevant boundary conditions. Once one provides a description of the physics
of the ocean, the solution of these equations should be straightforward.
Here we opted to discuss the effect that general relativity has on various 
classes of ocean modes in a less quantitative way. We simplified our equations
to the conditions that should prevail near the surface of a neutron star. This
allowed us to derive a dispersion relation which we used to 
deduce to what extent the various classes of modes are affected by the gravitational redshift 
and the rotational frame-dragging. Thus we have shown that the 
mode frequencies are typically lowered by roughly 20\%. This is not 
particularly surprising, but it should be noted that the overall effect 
is due to different combinations of the frame-dragging and the redshift
factors for the different kinds of modes. We also discussed (briefly) the 
effect that general relativity has on the ocean r-modes.
   
Our estimates provide a strong argument in favour of using the 
derived formulae
in more detailed (numerical) studies. In particular, we believe  
it is clear that one 
must  account for the relativistic effects if the aim is to identify 
the individual ocean modes present in observed data. 

\section*{Acknowledgements}

We acknowledge support from the EU Programme  
'Improving the Human Research Potential and the Socio-Economic Knowledge  
Base' (Research Training Network Contract HPRN-CT-2000-00137).  
In addition NA is grateful for support from the Leverhulme Trust via a Prize  
Fellowship, and would like to thank
the Institute for Theoretical Physics at the University  
of California - Santa Barbara for generous hospitality during the workshop  
``Spin and Magnetism in Young Neutron Stars'' where  this research  
was initiated. Among the participants of the programme, 
Greg Ushomirsky and Phil Arras are thanked for especially useful 
discussions.


\begin{thebibliography}{12}

\bibitem[\protect\citeauthoryear{Abramowicz, Rezzolla and Yoshida}{2000}]{rezzolla}
Abramowicz, M.A., Rezzolla, L., Yoshida, S., 2000, Class. Quantum, Grav \textbf{19}, 191
 
\bibitem[\protect\citeauthoryear{Bildsten and Cutler}{1995}]{bc}
Bildsten, L., Cutler, C., 1995, Astrophys. J., \textbf{449}, 800

\bibitem[\protect\citeauthoryear{Bildsten, Ushomirsky and Cutler}{1996}]{buc}
Bildsten, L., Ushomirsky, G., Cutler, C., 1996, Astrophys. J., \textbf{460}, 827

\bibitem[\protect\citeauthoryear{Hartle}{1967}]{hartle}
Hartle, J., 1967, Astrophys. J., \textbf{150}, 1005 

\bibitem[\protect\citeauthoryear{Heyl}{2001}]{heyl}
Heyl, J.S., preprint astro-ph/0108450

\bibitem[\protect\citeauthoryear{Ipser and Lindblom}{1992}]{ipser}
Ipser, J.,  Lindblom L., 1992, Astrophys. J., \textbf{389}, 392

\bibitem[\protect\citeauthoryear{Lee and Saio}{1997}]{ls}
Lee, U., and Saio, H., 1997, Astrophys. J., \textbf{491}, 839 

\bibitem[\protect\citeauthoryear{Lockitch, Friedman and Andersson}{2003}]{lock}
Lockitch, K.H., Friedman, J.L., Andersson, N., 2003, to appear in Phys. Rev D, preprint
gr-qc/0210102 

\bibitem[\protect\citeauthoryear{Miles}{1977}]{mil}
Miles, J., 1977, Proc. R. Soc. London, \textbf{353}, 377

\bibitem[\protect\citeauthoryear{M\"uller, Kelly and O' Brien}{1993}]{mkb}
M\"uller, D., Kelly, B.G., O' Brien, J.J., 1993, Phys. Rev. Let., \textbf{73}, 11

\bibitem[\protect\citeauthoryear{Provost, Berthomieu and Rocca}{1981}]{provost}
Provost, J., Berthomieu, G., Rocca, A., 1981, Astron. Astrophys., \textbf{94}, 126

\bibitem[\protect\citeauthoryear{Strohmayer and Bildsten}{2003}]{stroh}
Strohmayer, T.E., Bildsten, L.,  {\em New Views of Thermonuclear Bursts}
to appear in Compact Stellar X-Ray Sources, eds. W.H.G. Lewin and 
M. van der Klis, Cambridge University Press,
preprint astro-ph/0301544

\bibitem[\protect\citeauthoryear{Townsend}{2003}]{town}
Townsend, R.H.D,  2003, MNRAS, \textbf{340}, 1020

\end{thebibliography}
\end{document}